\documentclass[10pt]{article}
\usepackage{amsmath, amssymb, epsfig, multicol, indentfirst}
\begin{document}
\begin{center}
\Large
Exploring the Gauge Hierarchy Problem at $N = 2$ with TeV$^{-1}$-sized
Black Holes\\[0.25in]

\normalsize
Jason Minamora$^{1, 2}$\\[1em]
{\footnotesize $^{1}$\emph{California Institute of Technology, Department 
of Physics, Mailcode 356-48, 1200 E. California Blvd., Pasadena, CA 91125}\\
$^{2}$ \emph{Institute of Geophysics and Planetary Physics, University of 
	California, Los Angeles, Box 951567, Los Angeles, CA 90095}}
\end{center}
\small
\begin{abstract}
\noindent A mass Hierarchy of magnitude 
$O(10^{16})$ GeV separates the gravitational and electroweak interactions. 
Traditional proposed resolutions of this anomaly have included 
supersymmetric theories (most notably, the MSSM and ESSM) and string 
theory. However, energiers at which supersymmetric particles are expected to 
appear are not accessible to experiment, and string theory is only 
testable at $\sim \! M_{\text{pl}}$. A novel idea involves introducing 
extra dimensions into a Minkowski space-time to reduce the effective
separation between the gauge and gravitational couplings. Theoretical
developments of the so-called Arkani-Dinopolous-Dvali (ADD) 
extra-dimensional scenario, and its experimental verification, are presented
here.
\end{abstract}
\normalsize
%
%
\section{Background}

The gap between the strengths of the electromagnetic and gravitational interactions places constraints on the applicability of the Standard Model (SM). 
The cutoff scale between the gauge ($M_{\text{EW}} = 10^{3}$ GeV) and gravitational forces ($M_{\text{g}} \approx M_{\text{Pl}} = 10^{19}$ GeV) is 
large~\cite{one} ($\Lambda \sim 10^{16}$), suggesting a necessary 
extension to the SM.

Supersymmetric extensions to the SM have been proposed to resolve the gauge hierarchy. The Minimal Supersymmetric Standard Model (MSSM)---see, 
for example,~\cite{two} 
---assigns fermion superpartners to gauge bosons (including the unseen Higgs), and the spin-$\frac{3}{2}$ gravitino to the graviton. These particles, however, are yet to be seen; see Table 1, and also ref.~\cite{three} for discussion.

\begin{table}[h]
\begin{center}
\begin{tabular}{l l l}
\hline\hline
SM Particle(s)&MSSM Counterpart(s)&UL(MSSM) [GeV] (CL = 95\%)\\\hline
$\gamma, Z^{0}, H^{0}$&$\overset{\sim}{\chi}^{0}_{i}$ (neutralinos) &66.4\\
$W^{\pm}, H^{\pm}$&$\overset{\sim}{\chi}^{\pm}_{i}$ (charginos) &67.7\\
$e$&$\overset{\sim}{e}$     &95.0\\
$g$&$\overset{\sim}{\chi}_{g}$ (gravitino) &(no data)\\
\hline\hline\\
\end{tabular}
\caption{Upper limits on MSSM particle searches. Note that all values are 
within experimental reach. Values taken from ref.~\cite{four}.}
\end{center}
\end{table}
An alternative approach to the problem (discussed here) is to introduce extra spatial dimensions, each of radius $R$. In this scheme, the Planck scale is no longer fundamental. 
Instead, a test mass $m$ will experience a potential 
\begin{equation}
V(r) = \frac{m}{M^{N+2}_{*}R^{N}}\frac{1}{r} \quad (r \gg R)\label{E:N} \, 
\end{equation}
a distance $r$ from the source. $M_{\text{pl}}$ is now replaced with the 
effective ($4+N$)-dimensional Planck mass~\cite{five}:
\[
M^{2}_{\text{Pl}} = M^{2+N}_{*}R^{N} .
\]
If we choose $M_{*} \approx M_{\text{EW}}$, 
then gravity is of the same order as the gauge forces; substituting in 
figures gives
\[
R \approx 10^{\frac{30}{n} - 17} \times \left(\frac{1 \text{ TeV}}{m_{\text{EW}}}
\right)^{\frac{2}{n} + 1}.
\]
The distance scale for $N = 1$ is too large---$R$ = 1 AU. The $N = 2$ case,
which gives TeV$^{-1}$ (millimeter-) sized $R$, is empirically 
accessible~\cite{six}. 

One of the more interesting implications from the $N = 2$ theory 
is production (and decay) of millimeter-sized black holes (BHs) at 
experiments~\cite{seven, eight}. The simplest scenario arises when one considers 
Hawking-type BHs. From a classical, two-dimensional 
viewpoint~\cite{nine}, Hawking black holes decay forever. In (4 + 2) dimensions, 
the BH lifetime can be taken to 
be instantaneous. This is convenient, since one can produce 
and compare the 
decay spectrum of  
$N = 2$ BHs against the (higher-dimensional) Hawking spectrum in a short time.
If BH branching fractions agree with those predicted by the Hawking model,
a new upper limit could be placed on the quantum gravity scale.

\section{Phenomenological Considerations}

The simplest description of a Hawking BH is the semiclassical one, valid for
masses $M_{\text{BH}} \gg M_{*}$.\footnote{Here, $M_{*}$ is the $N = 2$ 
effective Planck mass.} 
The semiclassical approximation fails, however, when $M_{\text{BH}} \sim M_{*}$, since stringy corrections are no longer negligible. For our purposes, though, we will restrict discussion to the former case, where the spectroscopy of BH decays is well-described~\cite{ten}.

\subsubsection*{Production and Decay of TeV$^{-1}$ -sized Black Holes}

If BHs are produced with masses much larger than $M_{*}$, the 
Schwarzschild radius $R_{s}$ can be found using classical 
considerations~\cite{eleven}. For the six-dimensional case, 
\begin{equation}
R_{s} = \pi^{-1/3}\left(\frac{3 M_{\text{bh}}}{2 M_{*}}\right)^{\frac{1}{3}}
\frac{1}{M_{*}} \, .\label{E:sch}
\end{equation}
In particular, a reduced mass $M_{*}$ of 1 TeV places an upper limit on 
the size of these objects in the millimeter range, for production 
energies of up to $\sim 10$ TeV. These energies are presently above the thresholds of experiment, but the mechanism for production is still valid.

\begin{figure}[t]
\begin{center}
\begin{tabular}{l r}
\epsfig{figure = 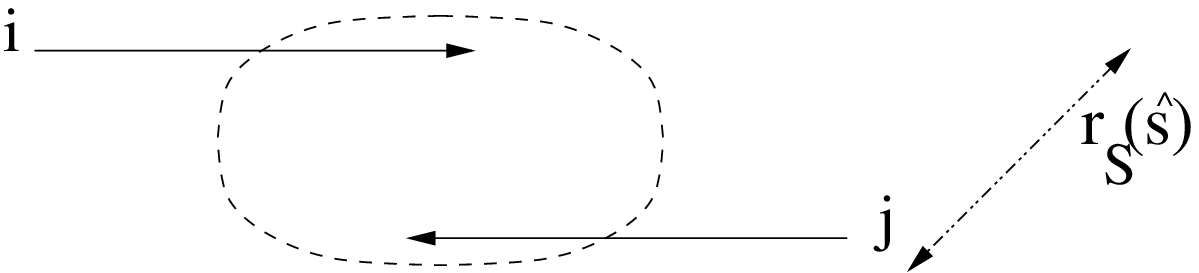,
width = 3.00in, height = 1.0in}
&
\epsfig{figure = 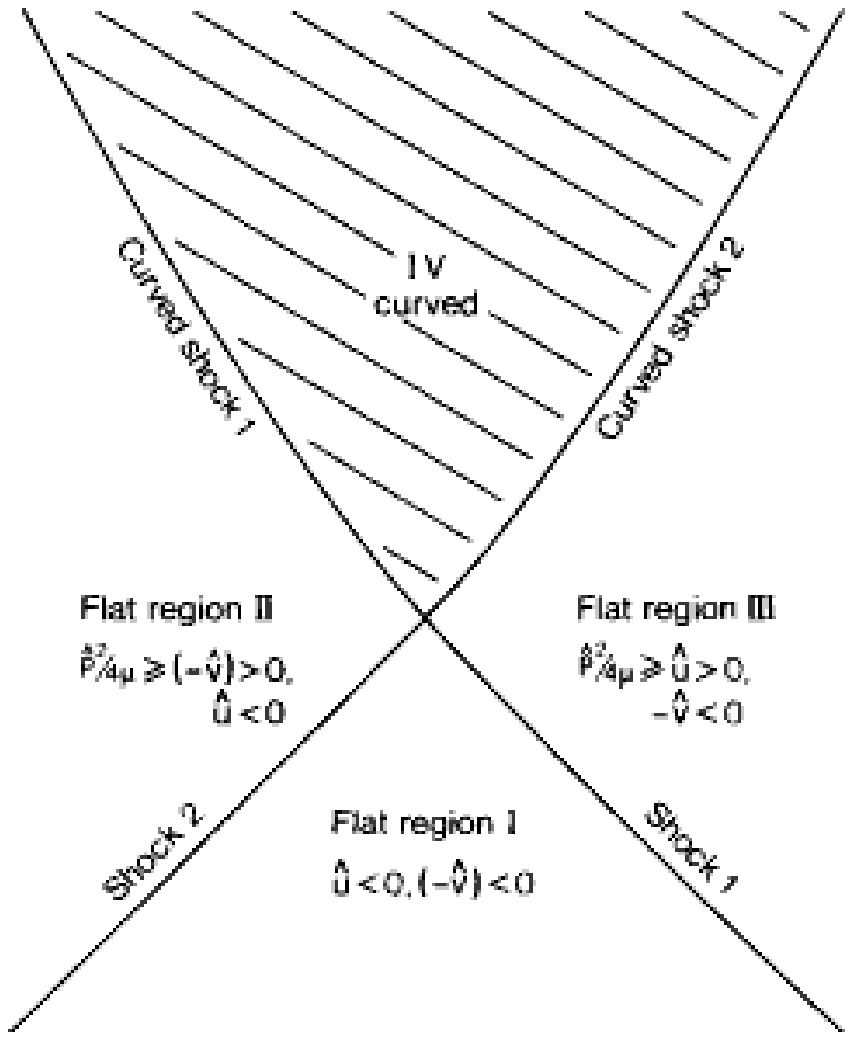,
width = 1.75in, height = 2.25in}
\end{tabular}
\small
\caption{Parton scattering. (\emph{1A---Left}) Close-range interactions
between two partons $i$ and $j$ produce a vortex of size $\sim r_{\text{S}}
(\hat{s} = E_{\text{cm}})$. (\emph{1B---Right}) The 
topological profile of space changes abruptly near the scattering region.
Regions (I), (II) and (III) fall outside the area cartooned in Fig. 1A; space
there is flat ($g_{\mu \nu}$ = $\eta_{\mu \nu}$).  In the production region 
(IV), the metric may be of Schwarzschild form, with a singularity at the 
center.}
\end{center}
\end{figure}

Consider the parton-parton\footnote{The six quarks and gluons.} 
scattering process shown in Fig. 1. Such reactions may take place at 
the LHC, and in collisions between neutrinos and baryons.  If the impact 
parameter $R_{s}(\hat{s})$ is smaller than $R_{s}$, 
we expect a vortex---here, a black hole---to form~\cite{eight}.


In our approximation, millimeter-sized black holes decay via the Hawking 
spectrum~\cite{twelve},
\begin{equation}
\frac{dN_{i}}{dt} \sim \frac{\Gamma_{s}(E, s)}{e^{E/T} - (-1)^{2s}}
\frac{d^{5}k}{(2\pi)^{5}}.\label{E:hawking}
\end{equation}
The \emph{greybody factor} $\Gamma_{s}$ is dependent on both the spin $s$, and the energy $E$; $\Gamma_{s}$, then, is effectively an absorption cross-section for particle species $i$. Because the Hawking distribution law~\eqref{E:hawking} contains a term in the (normalized) five-momentum, $d^{5} k$, we see that energy is radiated faster from a 
TeV$^{-1}$-sized black hole, than from a classical Schwarzschild black hole,
because of the increased phase-space volume available to the decaying 
particle.
In fact, lifetimes for the six-dimensional case are approximately 1 fs.

Expression~\eqref{E:hawking} bears resemblance to the Planck 
distribution function for a classical blackbody radiator:
\begin{equation}
 \frac{dN(\lambda, T)}{dE} \sim \frac{1}{(\lambda T)^{2}}\frac{1}{e^{1/\lambda kT}
\pm 1} \, .\label{E:class}
\end{equation}
The prefactor in $\lambda^{-2}$ indicates that~\eqref{E:class} is valid in the three-dimensional case (if we identify the de Broglie momentum $p = 1/\lambda$). Classical BHs are, therefore, longer-lived than those in six-dimensional space. 
Consequently, the momentum factor in~\eqref{E:hawking} 
suggests that millimeter-sized black holes decay into 
dominantly soft components (such as jets) compared 
to the macroscopic case~\eqref{E:class}.

One major obstacle is controlling Standard Model (SM) background. 
Because the interaction between partons is mediated by either a $\gamma, W^{\pm}$, or $Z^{\circ}$, there may be a contribution from the mode $Z^{\circ}
\rightarrow ee$ to the leptonic component of the spectrum.

\subsubsection*{Simulation of Black Hole Events}

The scattering process outlined in Fig. 1 is the 
primary means for producing black holes at the LHC and
elsewhere~\cite{seven, eight, twelve}; software to simulate the interaction has already been 
constructed~\cite{fourteen}. Current simulation techniques model parton evolution from $pp$ and $p\overline{p}$ processes---those most often studied at the LHC.

The software used to predict the decay products of 
black holes is not subject to the two major 
constraints of the theory outlined above; modern BH event generators  
include analyses of time-varying behavior---that is, temperature and spectroscopy---and attempts to describe the terminal state of BH decay. One shortcoming of the semiclassical approximation is a steady-state model that assumes equal branching fractions for all of the 60 particles in the SM~\cite{five}; that 
assumption is relaxed in a stochastic model, since the greybody 
factor $\Gamma_{s}$ is a free parameter in the simulation. Table 2 shows results from a Monte-Carlo run using the CHARYBDIS
analysis package~\cite{thirteen}.

\begin{table}
\begin{center}
\begin{tabular}{|c | c c c|}
\hline
&   &Energy [GeV] &\\
Species&Est. Mean ($\pm 50$ GeV)&&Est. 90\% UL ($\pm 50$ GeV)\\\hline\hline
$H^{0}$&500&&1250\\
$e^{\pm}$&500&&1250\\
$\gamma$&650&&1200\\\hline
\end{tabular}
\caption{Mean values and 90\% upper limits on energy 
for simulated decay spectra. 
Values taken from~\cite{thirteen}.}
\end{center}
\end{table}

\section{Experimental Foundations (Proposed)}

A host of ground-based and fixed-target experiments plan to begin 
searches for black holes once TeV physics is empirically realized. 
Among these are the 
LHC~\cite{seven, eight} (which will include the ATLAS 
detector~\cite{seventeen}), and   
AMANDA/ICECUBE~\cite{fourteen, fifteen, sixteen}.

\subsubsection*{Partonic scattering at the LHC}

Proton-anti-proton scattering with center-of-mass energy $E_{\text{cm}}$ 
greater than
the $N = 2$ BH production threshold $M_{\text{BH}}$ can be done with 
current experimental thresholds. 
experiment. If BHs are produced, they will be produced copiously at the 
LHC.

	Collider experiments have already established empirical lower limits
on center-of-mass production energies; the most recent summary reports
a 1.3-TeV threshold for the $N = 2$ case. This is below the phenomenological
limit derived from the higher-dimensional Newton's constant, however
($E_{N = 2} \gtrsim 1.6$ TeV), and so searches in higher-energy regimes
still look promising~\cite{eighteen}.

	The LHC collider beams are tentatively scheduled to operate at
7.0 TeV, with $\sim 1 \times 10^{14}$ collisions per second. The BH
production rate, $\Gamma_{\text{BH}}$, depends 
on $\sqrt{\hat{s}_{\text{beam}}}$, 
the collision rate, and a composite efficiency 
$\epsilon = \epsilon_{\text{b}} \times \epsilon_{\text{l}}$, based on 
beam injection parameters and predicted losses during operation.
Estimates using simulated protons at injection and during procession around 
the LHC storage ring give
efficiencies between $10^{-5}$ and $10^{-4}$~\cite{nineteen}; 
we therefore expect 
$\Gamma_{\text{BH}} (p \overline{p} 
\rightarrow BH) \lesssim 10^{10}$ s$^{-1}$; simulated
BH production cross-sections give similar predictions; see Fig. 2A.

\begin{figure}[t]
\begin{center}
\begin{tabular}{l r}
\epsfig{figure =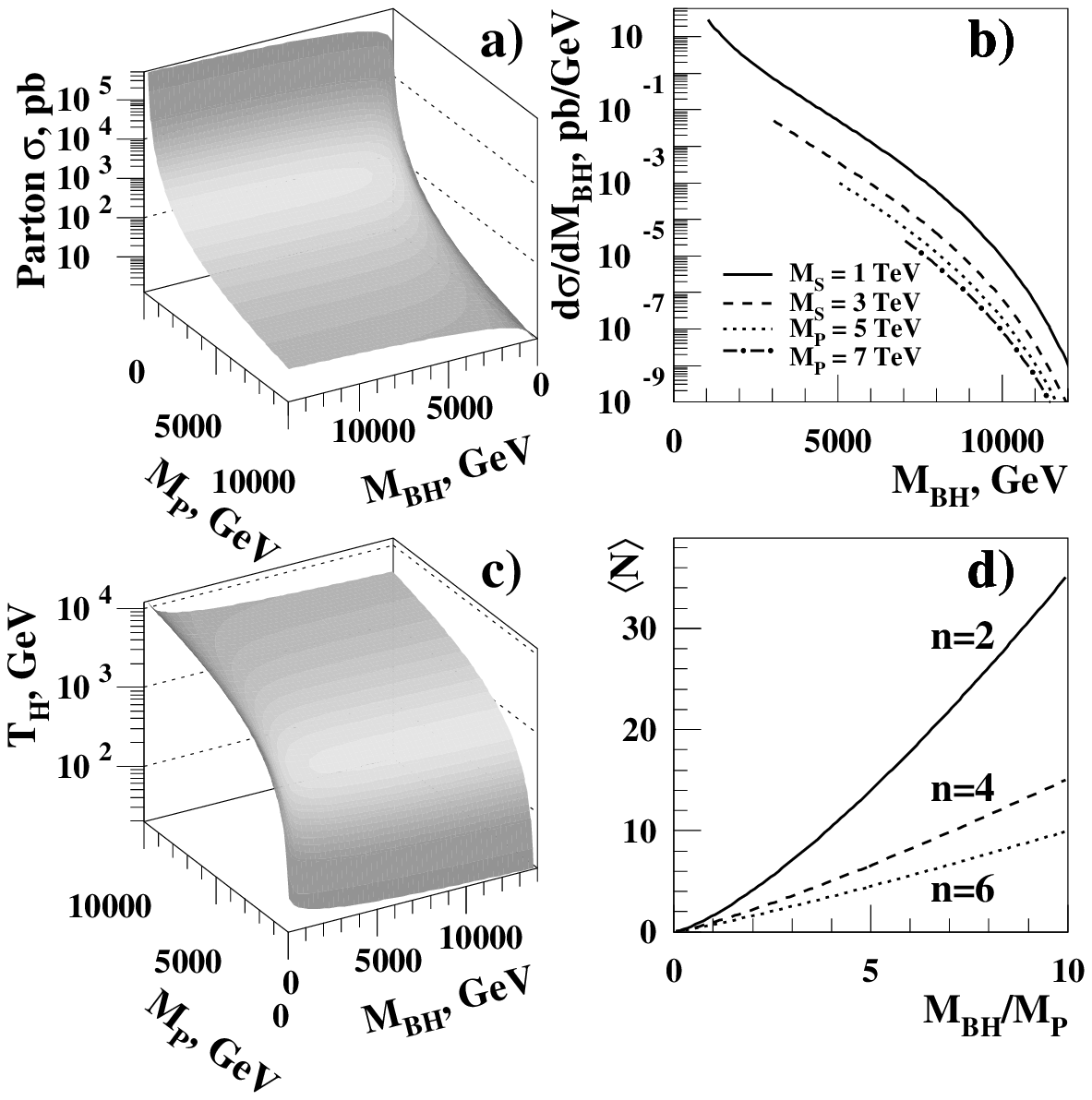,
width = 2in, height = 2in}
&
\epsfig{figure = 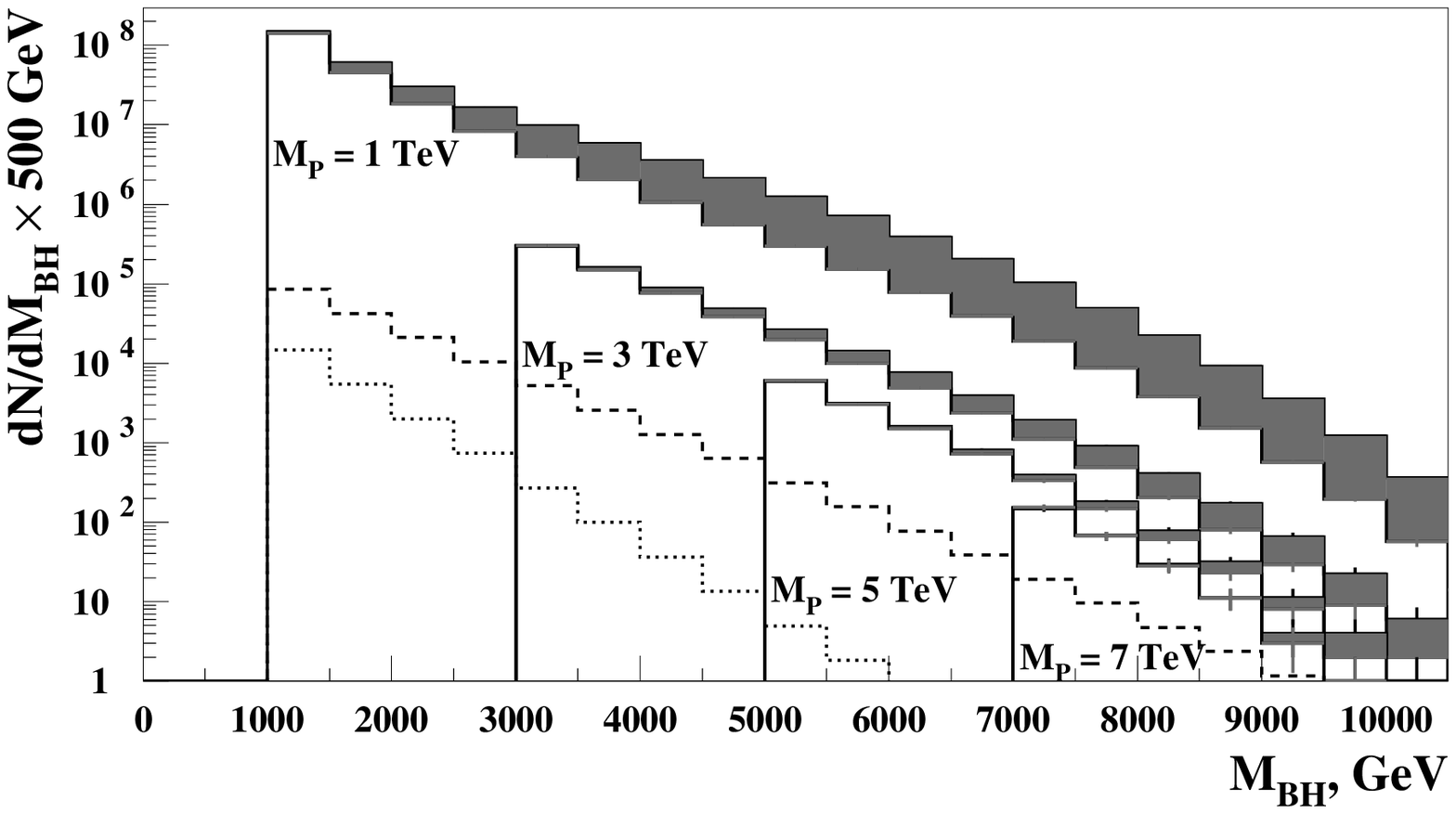,
width = 2in, height = 2in}
\end{tabular}
\small
\caption{(\emph{2A---Left}) Scale-dependent parton cross-sections 
[subpanel (a)]
and simulated multiplicity data  [subpanel (d)] LHC collisions
($\sigma_{\text{incl.}} = 1$ mb) should indeed produce BHs at a rate
$\Gamma_{\text{BH}}$ if they are seen, and are found to decay 
via the Hawking mechanism.  
(\emph{2B---Right}) Simulated number of BH events per unit reduced Planck mass 
, for four different values of $M_{\text{BH}}$. Shaded regions give ranges for
$2 \leq N \leq 7$. Dashed and dotted lines are expected contributions 
from the background modes 
$\Gamma (Z^{\circ} \rightarrow \ell^{\pm} \nu) + \Gamma (
Z^{\circ} \rightarrow \gamma + X)$, and $\Gamma (
Z^{\circ} \rightarrow \gamma + X)$ only, respectively~\cite{nine}.}
\end{center}
\end{figure}
\normalsize
	ATLAS, which uses the $p\overline{p}$ beam, plans to take its 
first science run in 2007. The proposed detector will be
sensitive to leptonic and electromagnetic decays from hard scattering events;
efficient detection of $\ell$ and $\gamma$ tracks 
is important for reconstructing BH decay products, since the 
observable portion of the Hawking spectrum is dominated by the intermediate 
vector bosons and charged leptons---see Table 4.
One must also be clever in separating SM background events due to the decay
of the $Z^{\circ}$, since products from leptonic decay channels may 
appear as BH signal.  This background can be significant, as shown in 
Fig. 2B.  Obtaining accurate counts from electromagnetic processes
also presents difficulty, because of large photon multiplicities 
from pseudoscalar meson decays---some examples of more exotic decays with 
significant electromagnetic contributions are shown in Table 3.
Placing kinematic cuts on mass 
differences between particles and decay products is a straightforward 
way to crudely correct for SM events. 

\begin{table}[h]
\begin{center}
\begin{tabular}{|l||l|l|}
\hline
\multicolumn{3}{|c|}{Selected $p \overline{p}$ Scattering Modes}\\
\hline\hline	&Primary Decay	&Final State\\
\hline\hline
$p \overline{p} \rightarrow 3\pi^{0}$	&$3 \pi^{0}$	&$6\gamma$\\\hline

	&$\pi^{0}\pi^{0}\eta^{\prime}$	&$10\gamma, 6\gamma$\\\cline{2-3}
$p \overline{p} \rightarrow 2\pi^{0} + X$	
	&$\pi^{0}\pi^{0}\omega$		&$7\gamma$\\\cline{2-3}
	&$\pi^{+}\pi^{-}\eta^{\prime}$	
		&$\pi^{+}\pi^{-1} 6\gamma$ (Dominant)\\\hline
	&$\pi^{0}\omega \eta$		&$7\gamma$\\\cline{2-3}n
$p \overline{p} \rightarrow \pi^{0} + X$	
	&$\pi^{0}\eta\eta$		&$6\gamma$\\\cline{2-3}
	&$\pi^{0}\eta\eta^{\prime}$	&$6\gamma$\\\hline
\end{tabular}
\caption{Electromagnetic $p \overline{p}$ channels with potential,
non-trivial contributions to the $\gamma$ component of the BH 
Hawking decay spectrum. Many such secondary decays are expected to be 
seen at the LHC; see ref.~\cite{twenty} for a full account.}
\end{center}
\end{table}

\subsubsection*{Cosmic-Ray Production of BHs}

	Ultra-high-energy photons from extragalactic sources 
produce neutrinos in matter, which are readily available for 
detection by ground-based detectors.
The parton scattering model discussed earlier can similarly be used 
here to discuss BH production in extensive air showers. Photons that initiate
electromagnetic cascades are typically in the range $10^{15}$--$10^{21}$ eV.
	For reasons to be discussed below, ground-based detectors have the
highest likelihood of seeing BH events, because of the low background. 
Neutrino arrays are sensitive
to the highest-energy portions of the cascade spectrum; BH formation should 
thus take place simultaneously with presently observed neutrino-matter
interactions. If BHs are to be produced via parton interactions, then 
the reaction $\nu N \rightarrow BH + \mu^{\pm}\ell^{\mp}X$ 
warrants consideration. Expected BH event rates (based on the ADD 
scenario~\cite{eleven}) for the South Pole-based ICECUBE neutrino array
have been calculated based on cosmic neutrino fluxes~\cite{seven, eleven}:
\begin{align}
\Gamma_{\text{BH}} = \int N(E) \frac{d \sigma}{d \Omega} d \Omega &= 
	n_{p} \times \int_{M_{*}}^{E^{\nu}_{\text{max}}}
	\frac{d\sigma}{dE} \Phi(E) \, dE\notag\\
	&= (3.5 \times 10^{-35}) N \quad \text{(events/yr)} \, .\label{E:tot}
\end{align}

	Here, $\sigma = \sigma(E)$ is the ADD-estimated cross-section for 
particle fluxes $\Phi(E)$. For $N = 10^{39}$, the number of 
protons in 1 km$^{3}$ of ice, $\sim$100 BHs per day 
should be seen from neutrino-nucleon scattering at ICECUBE.\footnote{The rate
calculated from~\eqref{E:tot} is for a six-dimensional BH (the $N = 2$ case).}

	Cosmic-ray experiments
offer a distinct advantage over beam-beam collision facilities like the LHC:
the freedom to choose between different active sources at various 
distances from Earth makes background correction a very simple task.
Figure 3 shows the signal-to-noise ratio plotted against kinematic and 
spatial parameters. Depending on the intensity of the source, background
begins to dominate at some critical distance $d_{c}$ (Fig. 3A). For objects 
at distances $d < d_{c}$, the effective area of air showers fully 
covers the field of view of the detector; background is no longer 
significant, allowing easy reconstruction of decay matter from BHs.
Fig. 3B takes the form of a ``rate-versus-threshold'' curve: signal is 
ultimately detector-limited, and is seen to level off at 
Lorentz factors $\gamma \sim 450$.
However, the shape of such a curve does not yield much scientific 
information, since it is the choice of threshold energy that determines 
photomultiplier tube response to cascades. 
The spectral profile of gamma-ray bursts (GRBs) 
can be directly extracted from Fig. 3C; low levels of signal from far away 
sources in higher-energy regions ($\alpha \gtrsim 2.4$) can be attributed
to attenuation effects; ICECUBE is sensitive to the hard component
of the spectrum for GRBs that are relatively nearby (dotted line in Fig. 3C).
Using ICECUBE as a nominal example of large-scale ground-based detectors, 
we find that such detectors' ability to observe sources at a continuum
of distance scales, and their freedom to discriminate at different
thresholds, results in background minimization superior to fixed-target and 
beam-beam experiments.
	So far, black holes have not yet been observed in the mode 
$\nu N \rightarrow \text{BH} + X$; upper 
limits are given elsewhere~\cite{twentyone}.
Should one be produced, scintillations from heavier decay products
($W^{\pm}\text{s}, \ell$) will be seen in the ice, even though
semi-classical arguments include
all SM particles in the BH decay spectrum. Unlike the LHC,
ground-based detection enjoys a high signal-to-noise ratio, and 
should therefore
be sensitive
to not only the quarks and leptons (see Table 4), but also 
the more exotic SM particles. It is likely that the first BH event will
be seen in cosmic-ray apparatuses.
\begin{table}[h]
\begin{center}
\begin{tabular}{l c c c c c c c}
	&$q$	&$\ell$	&$g$	&$W^{\pm}, Z$ &$\gamma$
 	&$G^{0}, G^{\pm}$ &$H$\\

\hline
Relative Contribution(\%)	&52.94	&17.65	&17.65	&6.62	&2.21	&2.21
				&0.74\\
\end{tabular}
\caption{Relative contributions of SM constituents to the Hawking 
Decay Spectrum. Values derived from multiplicity data given in~\cite{twelve}.}
\end{center}
\end{table}

\section{Summary and Conclusion}

	Higher-dimensional theories can be used to bridge the gap 
between the electroweak (gauge) interactions and gravity. We have considered
one such application that introduces extra TeV$^{-1}$ (mm) -sized dimensions
into space, which in turn lowers the fundamental Planck scale $M_{*}$ to
within experimental bounds. Observing Black holes in $N = 2$ extra dimensions 
would provide tangible evidence of quantum gravity at the TeV scale. 
Parton-parton scattering in $p \overline{p}$ collisions, and in 
$\nu N$ scattering events, are two proposed production methods.
If $N = 2$ black holes exist, one should see their copious production and 
decay at the LHC, and at a lower rate in neutrino arrays such as 
AMANDA/ICECUBE. Because of differences in resolution and background levels,
the LHC and neutrino telescopes have their own advantages---high event rates
are usually sacrificed for clean signal. As of this writing, no black
holes have been seen in neutrino arrays; studies at the LHC have yet to be
done.

\begin{figure}[t]
\begin{center}
\begin{tabular}{l c r}
\epsfig{figure =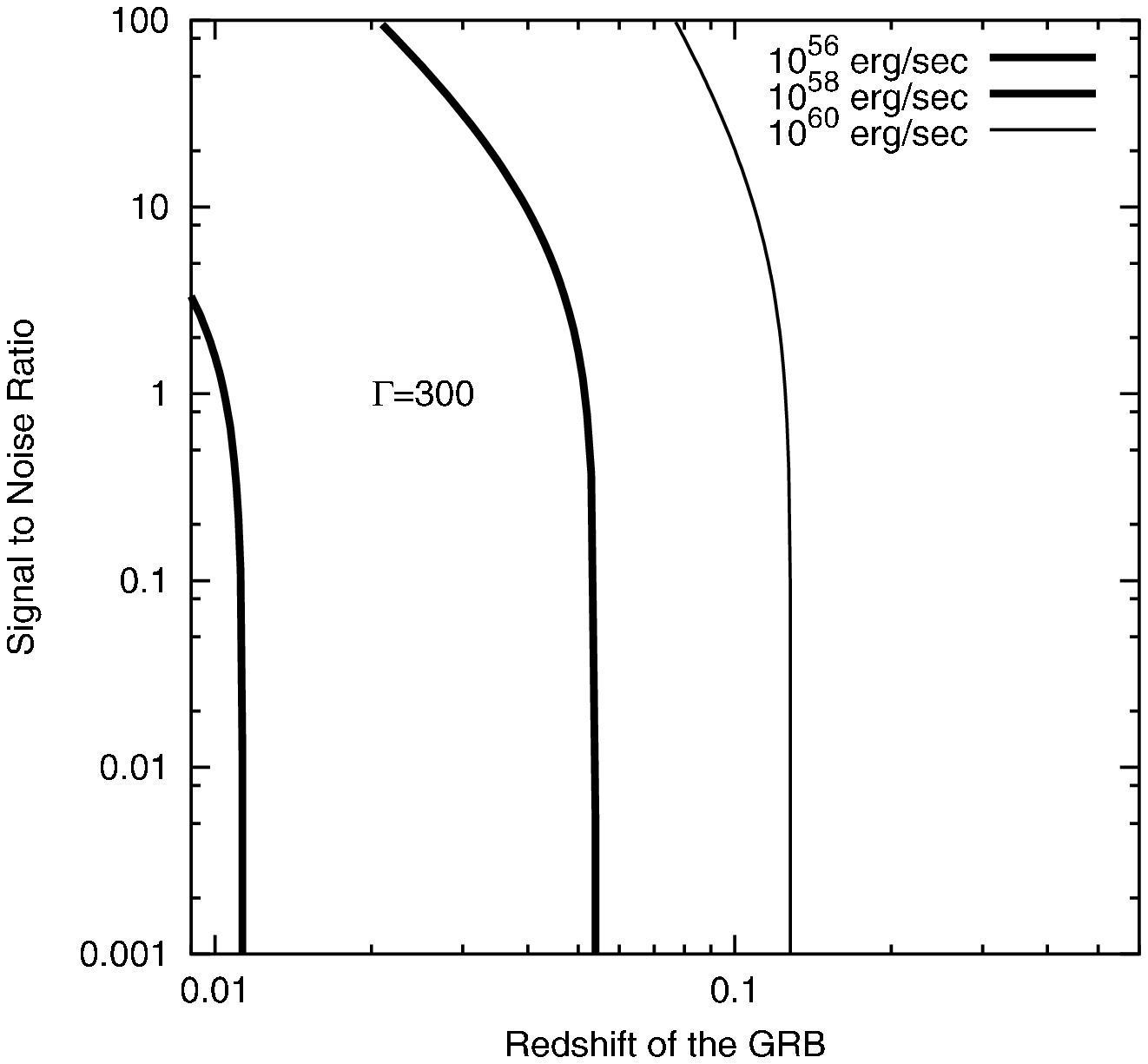,
width = 2in, height = 2in}
&
\epsfig{figure = 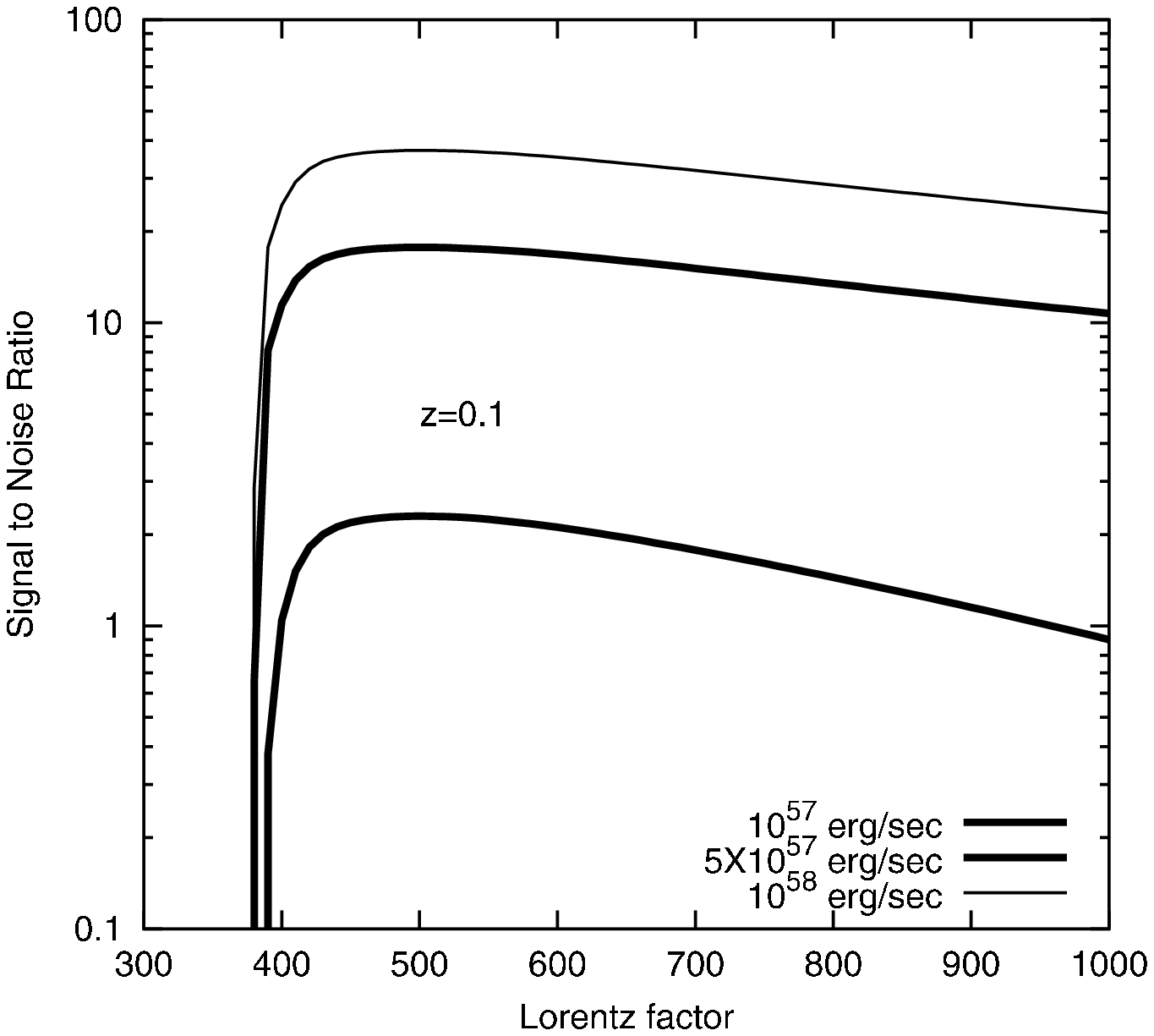,
width = 2in, height = 2in}
&
\epsfig{figure = 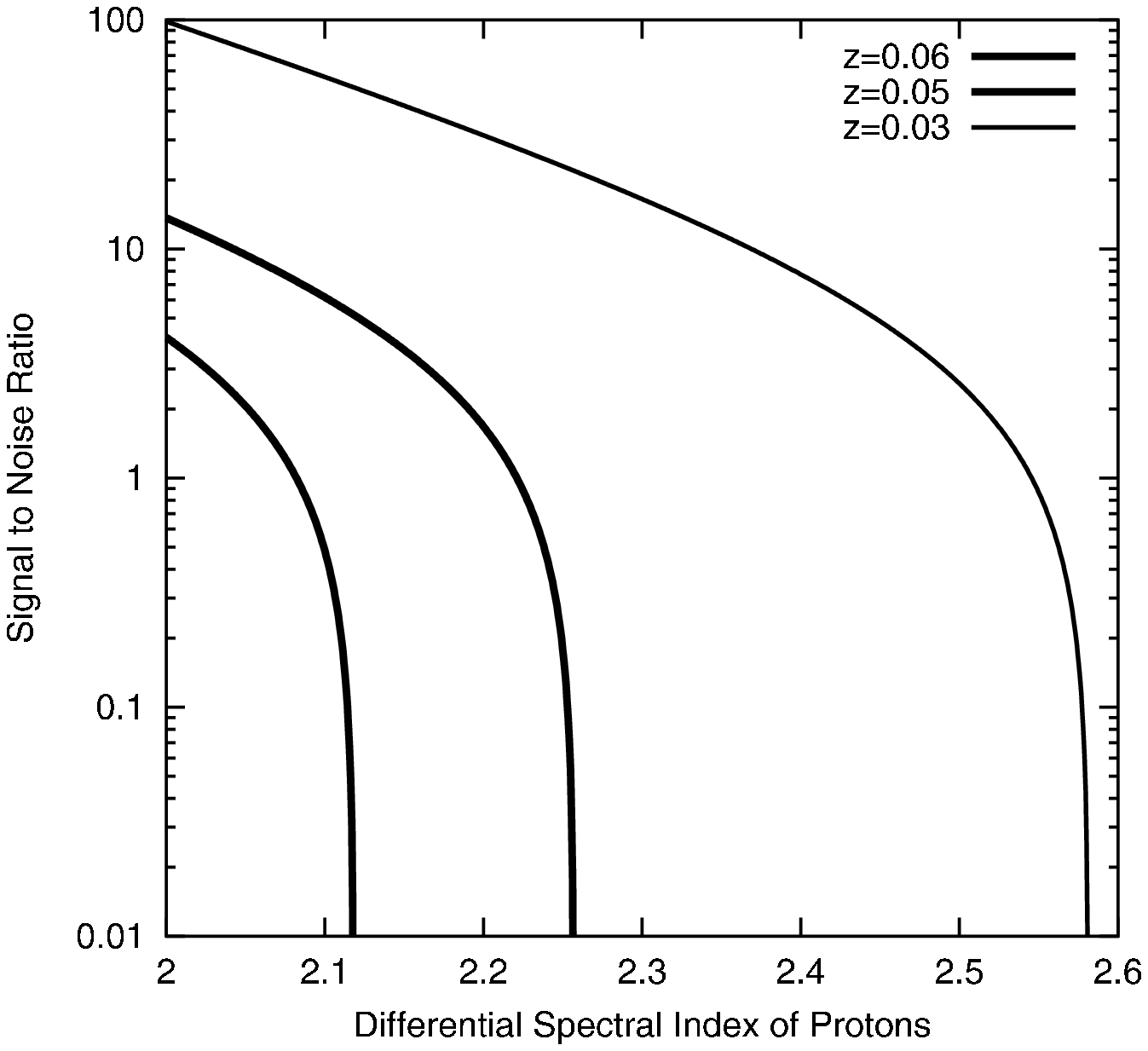,
width = 2in, height = 2in}
\end{tabular}
\small
\caption{Signal-to-noise ratio for ICECUBE-type detectors. Three regimes
are shown, for three different values of the luminosity---dependence 
on distance (\emph{3A, left}), outflow velocity (\emph{3B, center}), and
spectral index (\emph{3C, right})~\cite{sixteen}. These plots correspond to a
250 GeV threshold; normal incidence is assumed.}
\end{center}
\end{figure}
\bibliography{bh}
\bibliographystyle{unsrt}
\end{document}